\documentclass[%
 reprint,
 showkeys,
 amsmath,amssymb,
 aps,
floatfix,
]{revtex4-2}
\usepackage{graphicx}
\usepackage{siunitx}
\usepackage{amsmath}
\usepackage{braket}
\usepackage{dcolumn}
\usepackage{bm}
\usepackage{hyperref}

\newcommand{\Singlet}{${}^{1}\text{P}_1 \leftarrow {}^{1}\text{S}_0$}
\newcommand{\SingletWL}{$229$\,nm}

\newcommand{\Triplet}{${}^{3}\text{P}_1 \leftarrow {}^{1}\text{S}_0$}
\newcommand{\TripletWL}{$326$\,nm}

\newcommand{\cryotemp}{3\,K}

\newcommand{\GammaHalf}{\frac{\Gamma}{2}}
\newcommand{\GammaMHz}{$\Gamma/(2\pi)$}
\newcommand{\DeltaA}{\Delta_{1/2}}
\newcommand{\DeltaB}{\Delta_{3/2}}
\begin{document}

\preprint{APS/123-QED}

\title{High-resolution isotope-shift spectroscopy of Cd I}


\author{Simon Hofsäss$^1$, J. Eduardo Padilla-Castillo$^1$, Sid C. Wright$^1$, Sebastian Kray$^1$, Russell Thomas$^1$, Boris G. Sartakov$^1$, Ben Ohayon$^2$, Gerard Meijer$^1$ $\&$ Stefan Truppe$^1$}\altaffiliation[]{Current address: Centre for Cold Matter, Blackett Laboratory, Imperial College London, London SW7 2AZ.}
\affiliation{%
$^1$Fritz-Haber-Institut der Max-Planck-Gesellschaft, Faradayweg 4-6, 14195 Berlin, Germany\\
$^2$Institute for Particle Physics and Astrophysics, ETH Z\"urich, 8093 Z\"urich, Switzerland}

\date{\today}

\begin{abstract}
We present absolute frequency measurements of the \Singlet{} (\SingletWL{}) and \Triplet{} (\TripletWL{}) transitions for all naturally occurring isotopes of cadmium. The isotope shifts and hyperfine intervals of the fermionic isotopes are determined with an accuracy of 3.3\,MHz.
We find that quantum interference in the laser-induced fluorescence spectra of the \Singlet{} transition causes an error of up to 29(5)\,MHz in determining the hyperfine splitting, when not accounted for with an appropriate model.
Using a King-plot analysis, we extract the field- and mass-shift parameters and determine nuclear charge radius differences for the fermions. The lifetime of the $^1\text{P}_1$ state is determined to be 1.60(5)\,ns by measuring the natural linewidth of the \Singlet{} transition. 
These results resolve significant discrepancies among previous measurements. 
\end{abstract}

\keywords{isotope shift, quantum interference, radiative lifetime}
\maketitle


\section{\label{sec:introduction}Introduction}
The energy differences between isotopes of an atom or molecule are called isotope shifts (ISs). In atoms, the change in energy has two main contributions: the mass shift (MS) and the field shift (FS).
The mass shift is caused by changes in the electronic wavefunction upon altering the nuclear mass, whereas the field shift arises from changes in the nuclear charge distribution~\cite{King1963}. The shifts in the energy levels can be probed spectroscopically and if the ISs are caused by the MS and FS only, there is a linear relationship between the ISs of two transitions, known as the King plot linearity.
Small deviations from this linearity can be a sensitive probe for higher-order terms in the mass shift, the quadratic field shift or isotope-dependent nuclear deformation. 
In addition, precise values for the MS and FS factors, and deviations from the expected linear behavior of the King plot, provide a useful benchmark for atomic structure calculations.

Recently, it has been suggested that non-linearities in a King plot can arise from physics beyond the Standard Model (BSM) of particle physics~\cite{Delaunay2017, Frugiuele2017, Berengut2018}. A new intra-atomic force between a neutron and an electron, mediated by a new boson, can lead to an isotope-dependent energy shift and the introduction of a Yukawa-type particle shift results in a non-linear King plot~\cite{Ono2022}. However, this method of searching for new physics relies on a detailed knowledge of the Standard Model contributions. The Cd atom has recently attracted attention as a sensitive probe for new physics because of its six even-even isotopes (even number of protons and neutrons) that have a high natural abundance~\cite{Ohayon2022}. In addition, the Cd nucleus ($Z=48$) is only one proton pair below the $Z=50$ proton shell closure. This significantly reduces potential non-linearities that arise from a deformed nucleus, which currently limits the interpretation of isotope-shift measurements with Yb\cite{Allehabi2021, Hur2022}. Cd possesses a strong cooling transition and weak intercombination lines that can be used for narrow-line cooling, precision spectroscopy, and metrology~\cite{Gibble2018, Yamaguchi2019}, ideal for a sensitive search for BSM physics. 

We recently showed that combining precise isotope-shift spectroscopy with new, state-of-the-art atomic structure calculations, allows determining the differences in the radii of the nuclear charge distribution with high accuracy~\cite{Ohayon2022}. This provides an alternative, independent method to muonic X-ray spectroscopy or electron scattering. The charge radius is a fundamental property of the atomic nucleus, and precise measurements of small differences between isotopes through optical spectroscopy provide stringent tests for nuclear theory~\cite{Hammen2018, Koszorus2021}. In addition, highly accurate charge radii differences are critical to understanding the nuclear contributions to non-linearities in a King plot. 

Here, we present the spectroscopic method used in~\cite{Ohayon2022} to determine ISs of the bosonic \Singlet{} and \Triplet{} transitions in Cd I and combine our previous results with new measurements of the fermionic isotopes $^{111}$Cd and $^{113}$Cd. For the \Singlet{} transition, we use enriched Cd ablation targets and a polarization-sensitive detection scheme to assign spectral lines of different isotopes that otherwise overlap. We measure the hyperfine intervals in the $^1\text{P}_1$ state of the two stable fermionic isotopes $^{111,113}$Cd with MHz accuracy by analyzing subtle quantum interference effects in the laser-induced fluorescence. Knowledge of the exact lineshape allows us to significantly improve the ISs and resolve significant discrepancies among previous measurements. The radiative lifetime of the $^1\text{P}_1$ state is extracted by fitting the spectral lineshape. The absolute transition frequencies are determined with high accuracy. A King-plot analysis of the two transitions allows extracting the intercept and slope and to determine precise values for the differences in the nuclear charge radii of the fermions. This measurement is also used to benchmark a recent high-level atomic structure calculation of the MS and FS, with which we find excellent agreement~\cite{Schelfhout2022}. In addition, we show that the off-diagonal, second-order hyperfine interaction in the fermions is $\lessapprox 3$\,MHz, in good agreement with calculations~\cite{Schelfhout2022}. 

The methods presented here are relevant to the field of collinear laser spectroscopy of rare isotopes produced at accelerator facilities\cite{Garcia2016}. In these experiments, laser spectroscopy is used to determine the fundamental properties of nuclei, including the nuclear spin, the magnetic dipole moment, the electric quadrupole moment, and the charge radius. Due to the low number of atoms produced in these experiments, these properties are obtained from strong transitions in the visible or UV part of the spectrum to increase the signal-to-noise ratio. We show that quantum interference in the laser-induced fluorescence of strong transitions can cause significant systematic errors in determining the fundamental properties of nuclei.

\section{\label{sec:experimentalsetup}Experimental setup}
\begin{figure}
    \centering
    \includegraphics[]{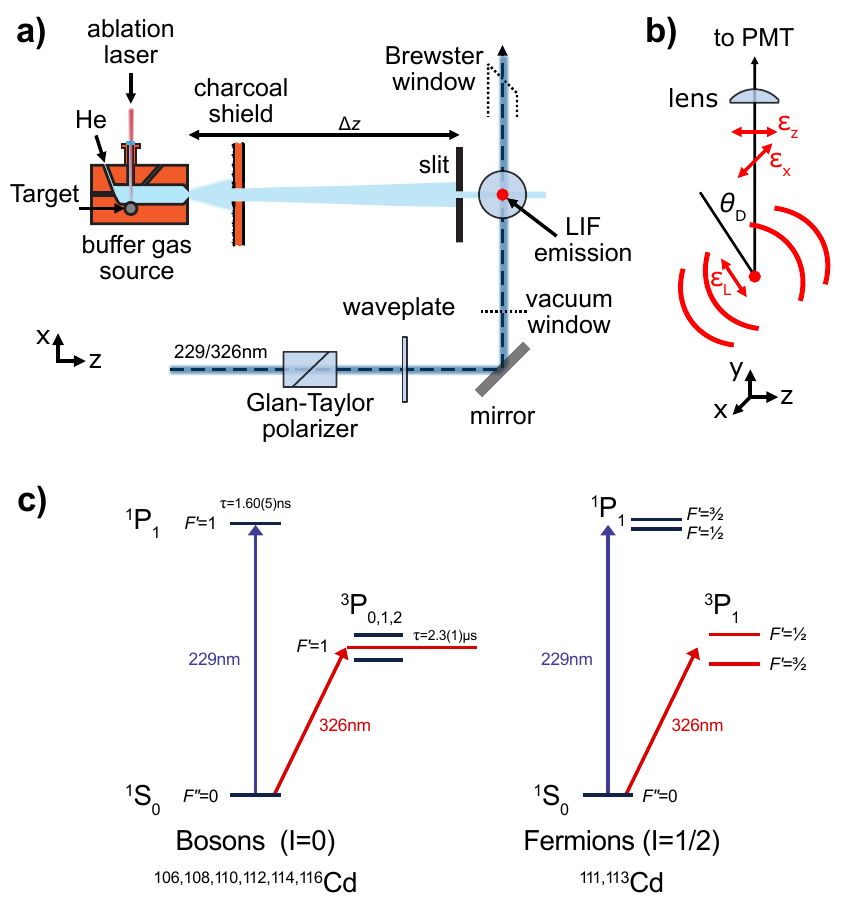}
    \caption{a) A collimated atomic beam from a pulsed cryogenic buffer gas source crosses the beam of a continuous wave UV laser at $\Delta z=0.73$\,m. The laser polarization is cleaned-up with a Glan-Taylor polarizer and is varied using a waveplate. b) Laser-induced fluorescence (LIF) is detected with a photomultiplier tube (PMT). The linear laser polarization ($\epsilon_L$) forms an angle $\theta_D$ with the detector axis. Emitted photons are polarized along $\epsilon_{x,z}$. c) Energy level diagram for naturally abundant cadmium isotopes.}
    \label{fig:experimental_setup}
\end{figure}
Figure \ref{fig:experimental_setup}a) shows a schematic representation of the experimental setup. 
We use a cryogenic buffer gas beam source to produce a slow, pulsed beam of Cd atoms with a brightness of about $10^{13}$ atoms per steradian per pulse and a forward velocity of $100-150$\,m/s\cite{Truppe2018, Wright2022}. A 1064\,nm beam from a pulsed Nd:YAG laser is focused onto a solid Cd metal target and creates a hot cloud of atoms by laser ablation. We use a multi-sample target holder, enabling fast switching between targets with different isotopic compositions. The vaporized atoms are cooled to \cryotemp{} by a continuous flow of 1\,sccm (standard cubic centimeter per minute) cryogenic helium buffer gas and are extracted into a beam through a 4\,mm aperture in the buffer gas cell. Charcoal-coated copper shields act as a sorption pump for the buffer gas to keep the pressure in the low $10^{-7}$\,mbar range. The atomic beam is probed in a laser-induced fluorescence (LIF) detector located 0.73\,m from the source aperture. A slit with a width of 2\,mm along $x$ restricts the transverse velocities of the atomic beam entering the LIF detector. This reduces the Doppler broadening of the \Singlet{} transition to below 2.7\,MHz for a forward velocity of 150\,m/s. To excite the \Singlet{} transition, we use a frequency-quadrupled continuous-wave titanium-sapphire (Ti:Sa) laser that can generate up to 200\,mW at \SingletWL{}. For the \Triplet{} transition at \TripletWL{}, we use a frequency-doubled continuous wave ring-dye laser (Sirah Matisse 2DX) with a frequency doubling module (Spectra Physics; Wavetrain) and a Pound-Drever-Hall locking scheme. This laser is stabilized to a linewidth of 100\,kHz using a temperature-stabilized reference cavity. The maximum output power is 80\,mW in the UV. The laser polarization is purified with an $\alpha$-BBO-Glan-Taylor polarizer ($1:10^5$ extinction ratio). The angle of the linear laser polarization with respect to the detector axis, $\theta_\text{D}$, can be adjusted with a $\lambda/2$ waveplate. LIF of atoms that pass through the detection zone is detected with a photomultiplier tube (PMT). The laser beam exits the chamber through a Brewster window to avoid back reflection into the interaction zone.

For the \Singlet{} transition, we use a laser beam with a diameter of 5\,mm, nearly flat-top intensity distribution, and a laser power of 0.5\,mW. This corresponds to a saturation parameter of $s_0=I_{\text{peak}}/I_{\text{sat}} \approx 1/400$, where $I_{\text{sat}}=\pi h c \Gamma/(3 \lambda^3)=1.1$\,W\,cm$^{-2}$ is the two-level saturation intensity and $\Gamma=1/\tau$, with $\tau=1.60(5)$\,ns being the excited-state lifetime (see below). The scattering rate for small $s_0$ on resonance is approximately $s_0 \Gamma/2 \approx 0.79\,$(µs)$^{-1}$. The mean interaction time of the atoms with the laser beam is about 30\,µs so that each atom scatters on average 24 photons. This results in a radiation-pressure-induced Doppler shift of 1.7\,MHz. In relative measurements, such a shift is smaller than our statistical error; for absolute measurements, it is negligible compared to the absolute uncertainty of the wavemeter. For the \Triplet{} transition, $1$\,mW of laser power in a Gaussian beam with a spot size of 10\,mm ($s_0\approx 10$) is sufficient to saturate the transition. The Doppler broadening due to the transverse velocity distribution in the detector is 1.5\,MHz and the radiation pressure detuning due to the scattering of 10 photons is 0.33\,MHz. 

The laser wavelengths are measured with a wavemeter (HighFinesse WS8-10) that is referenced to a calibrated, frequency-stabilized HeNe laser at 633\,nm and has a resolution of 0.4\,MHz. For the \SingletWL{} transition, we measure the frequency-doubled wavelength near 458\,nm, whereas for the \TripletWL{} we measure the fundamental wavelength near 652\,nm. 

\section{\label{sec:experiment}Atomic beam spectroscopy}
This section is split into three parts. In Section A we show measured spectra of the \Singlet{} and \Triplet{} transitions and explain the experimental methods and data analysis used. We benchmark a sophisticated model for the lineshape of the fermions and measure the linewidth of the \Singlet{} transition to infer the lifetime of the $^1$P$_1$ state. In Section B we discuss the results, analyze them in a King plot and calculate the nuclear charge radius differences for the fermions. Section C focuses on systematic uncertainty, which is the limiting factor in the accuracy of our measurements. We compare our measurements in Cd with known properties of the hyperfine intervals of the $^3$P$_1$ state and measure well-known transitions in atomic copper at nearby wavelengths. Finally, the accuracy of relative measurements is established by probing the linearity of our wavemeter with an ultra-stable cavity.
\subsection{Measurements}
\begin{figure}
    \centering
    \includegraphics[]{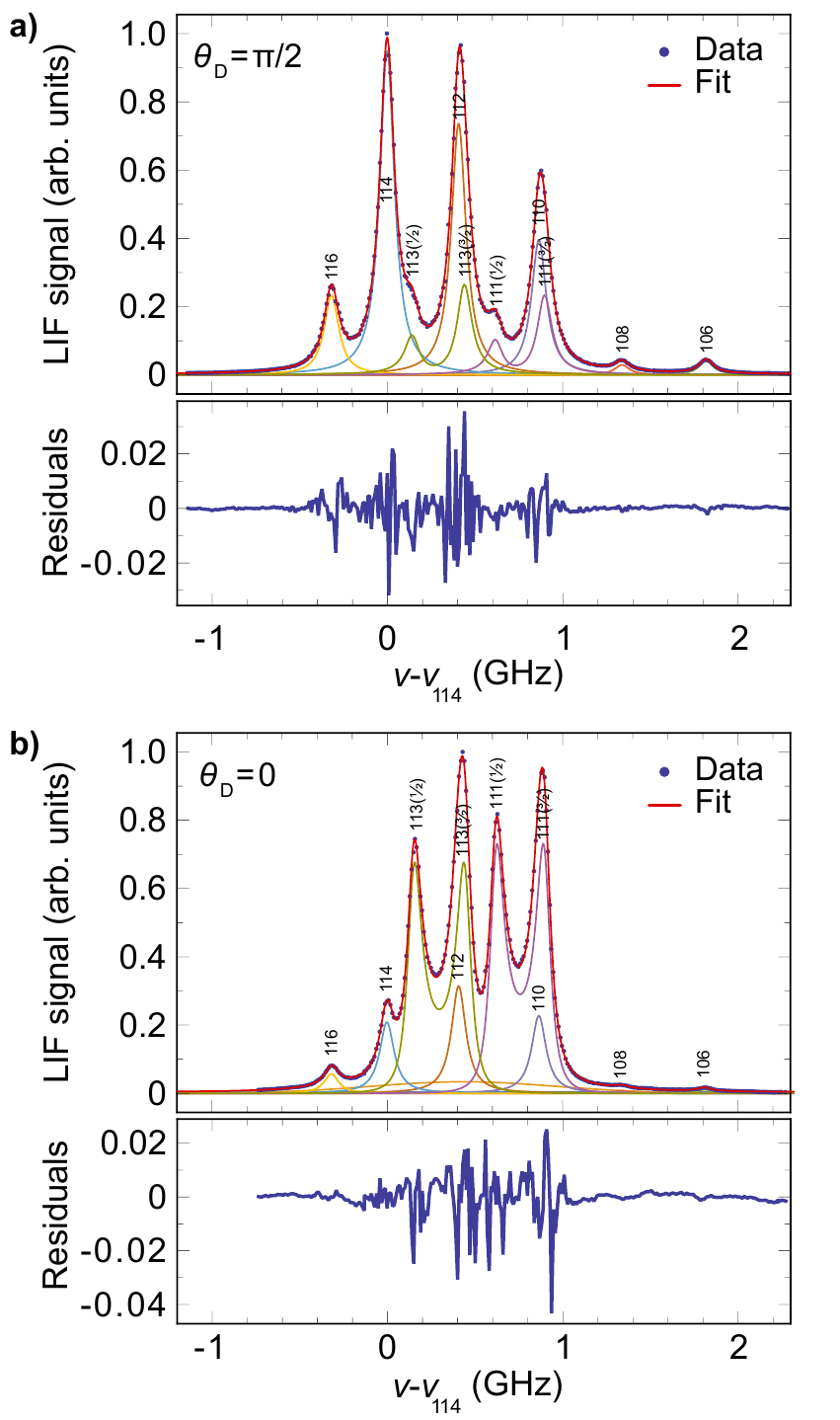}
    \caption{\label{fig:spectra}Isotope-shift spectrum of the \Singlet{} transition at \SingletWL{} relative to ${}^{114}$Cd, measured under two different laser polarizations angles, fitted with the quantum interference model for fermions. The fit residuals are shown below the respective plot. a) $\theta_D=\pi/2$ maximizes the laser-induced fluorescence emission of the bosons towards the detector. b) $\theta_D=0$ suppresses the fluorescence emission of the bosons and thus improves the accuracy in determining the transition frequencies of the fermions.}
\end{figure}
Figure \ref{fig:spectra} shows two fluorescence spectra around \SingletWL{} for two different laser polarization angles $\theta_D$. The natural linewidth of the \Singlet{} transition and the hyperfine splitting of the fermionic isotopes ($\Delta_\text{HF}$) in Cd are of the same order of magnitude as the ISs. The result is a spectrum with a significant overlap of the spectral lines. This overlap complicates the determination of the resonance frequencies, making them dependent on the precise determination of the lineshape. The spectroscopic lineshape of bosons and fermions is inherently different. Bosons, in contrast to fermions, have no nuclear spin and thus do not exhibit a hyperfine structure (see Figure \ref{fig:experimental_setup}c)). In this particular system of ($F=1/2$, $F'=1/2,3/2$), we see a strong influence of quantum interference for fermions. To separate between fermionic and bosonic peaks, we use isotope-enriched targets and take advantage of the differences between the fluorescence emission patterns of bosons and fermions.\\
\begin{figure}
    \centering
    \includegraphics[]{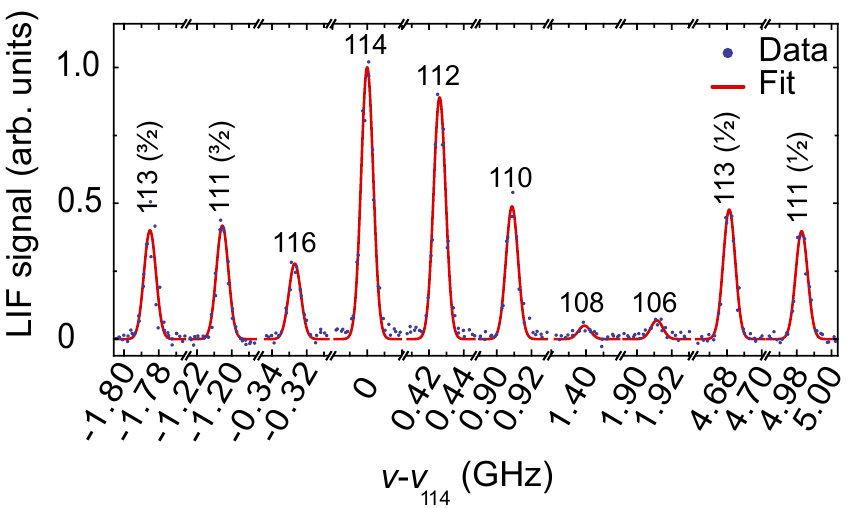}
    \caption{Representative spectrum for \TripletWL{} relative to ${}^{114}$Cd. The excited state hyperfine level of fermionic isotopes is indicated in brackets. 
    }
    \label{fig:326-spectrum}
\end{figure}
A typical spectrum for \TripletWL{} is shown in Figure \ref{fig:326-spectrum}. The lineshape is dominated by Gaussian broadening with a full-width at half maximum of 4.1\,MHz. There is no spectral overlap between the lines and the excited state hyperfine levels of the fermionic isotopes are split by approximately $10^5\Gamma$ with negligible influence of quantum interference. When the laser beam is not orthogonal to the atomic beam, the spectrum of atoms with a high forward velocity is shifted with respect to atoms with a low forward velocity. By changing the alignment of the laser beam to overlap the spectra, we reduce the residual Doppler shift to $\leq 1$\,MHz. We fit a Gaussian function to the spectral lines and determine the line centers with a statistical uncertainty of better than 1\,MHz. The results are summarized in Table \ref{tab:shift-comp}.

\subsubsection{Spectral lineshape}
\label{sec:lineshape}
The fluorescence emission pattern of the bosons corresponds to that of a classical Hertzian dipole. When the atoms are excited with linearly polarized light and the detection direction forms an angle $\theta_{\text{D}}$ to the polarization axis, the detected intensity is proportional to $\sin^2\theta_{\text{D}}$. Thus, when detecting in a small solid angle $\theta_{\text{C}}$ around $\theta_{\text{D}} =\{0,\pi\}$, the signal from the bosons is maximally suppressed. Following \citeauthor{Brown2013}~\cite{Brown2013}, the emission pattern for the bosonic species, $\mathcal{S}^{(b)}$, is given by

\begin{equation}
\begin{gathered}
\mathcal{S}^{(b)} = \frac{2}{3}\left(1- P_2(\cos\theta_{\text{D}}) g(\theta_{\text{C}}) \right) \frac{1}{(\Gamma/2)^2 + (\omega - \omega_0)^2} \hspace{0.2cm}, \\
\end{gathered}
\label{eq:lineshapeBosons}
\end{equation} 

\noindent Here, $\Gamma$ is the spontaneous decay rate of the excited state, $\omega$ is the angular frequency of the laser, $\omega_0$ is the transition angular frequency, $P_2(\cos\theta_{\text{D}}) = \frac{1}{2}(3 \cos^2\theta_{\text{D}} -1)$ is the second Legendre polynomial, and the factor $g(\theta_{\text{C}}) = \cos(\theta_{\text{C}})\cos^2(\theta_{\text{C}}/2)$ accounts for the finite solid angle (half-angle $\theta_{\text{C}}$) of the collection optics. In our measurements $\theta_{\text{C}}\approx0.11$.

\begin{figure}
    \centering
    \includegraphics[]{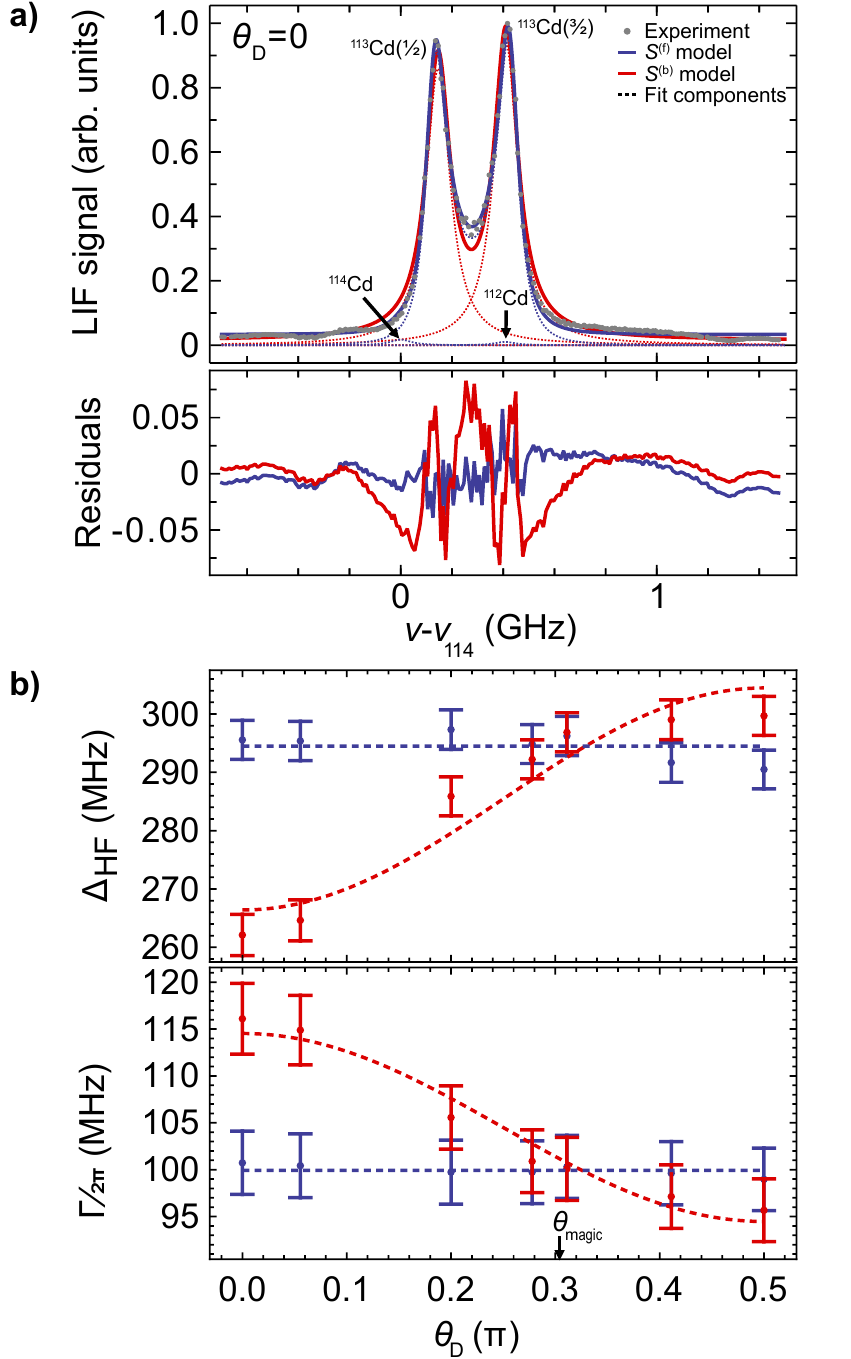}
    \caption{\label{fig:qint-experiment}Isolated hyperfine structure of the ${}^{113}$Cd isotope, measured using an enriched target at \SingletWL{}, plotted including the systematic error of $3.3$\,MHz. The quantum interference model ($\mathcal{S}^{(f)}$, blue) is compared to a sum of two Lorentzian line-shapes ($\mathcal{S}^{(b)}$, red). (a) The fit residuals reduce significantly by including quantum interference. (b) Measured hyperfine intervals $\Delta_\text{HF}$ and homogeneous linewidth $\Gamma/(2\pi)$ as a function of the polarization angle for the two lineshape models. Blue curves are our final result. The red dashed curve is a fit to the data.}
\end{figure}

The presence of hyperfine structure in the excited states of the fermionic isotopes has two effects. First, it changes the fluorescence emission pattern in comparison to the bosons. Upon excitation with linearly polarized light, the $F'=1/2$ excited magnetic sub-levels emit circular and linear light, which in sum is emitted isotropically, whereas the $F'=3/2$ sub-levels show anisotropic emission. 
Second, the hyperfine interval in the excited state is about $3\Gamma$, leading to significant interference between scattering paths. This alters the observed lineshape in fluorescence and can be seen as the time-averaged analog of quantum beats. To model the lineshape, we again follow \citeauthor{Brown2013}~\cite{Brown2013}. The fluorescence spectrum for the fermions, ${\cal S}^{(f)}$ is given by, 
\begin{equation}
\begin{gathered}
\mathcal{S}^{(f)} = A + (B+C) P_2(\cos\theta_{\text{D}}) g(\theta_{\text{C}}), \\
A = \frac{1}{(\Gamma/2)^2+ \DeltaA^2} + \frac{2}{(\Gamma/2)^2 + \DeltaB^2} , \\
B = -\frac{1}{(\Gamma/2)^2 + \DeltaB^2}, \\
C = -\left(\frac{1}{(\Gamma/2)^2 + \DeltaA\DeltaB + i \GammaHalf(\DeltaA-\DeltaB)} + c.c. \right) \hspace{0.2cm},
\end{gathered}
\label{eq:lineshapeFermions}
\end{equation} 
\noindent where $\Delta_{F'} = \omega - \omega_{F'}$. $\omega_{F'}$ denotes the transition angular frequency to the excited-state hyperfine component $F'$, and $c.c.$ is the complex conjugate. The first term of equation \eqref{eq:lineshapeFermions}, $A$, is the emission averaged over the total solid angle, represented by a sum of Lorentzians. The second term, $B$, represents the anisotropy of emission from the $F'=3/2$ excited state. The last term, $C$, accounts for the interference between decay paths. These last two terms reduce to zero by setting $\theta_{\text{D}}=\theta_{\text{magic}} = \arccos(\frac{1}{\sqrt{3}})$, the so-called ``magic angle". To illustrate this effect, we take spectra at seven different polarization angles, $\theta_D$, and compare the two models $\mathcal{S}^{(f)}$ and $\mathcal{S}^{(b)}$ to fit the observed line shapes. Figure~\ref{fig:qint-experiment}a shows a spectrum for $\theta_{\text{D}}=0$ using an enriched target (${}^{113}$Cd). Including the interference term, i.e., using $\mathcal{S}^{(f)}$, reduces the fit residual RMS by a factor of two. This is further substantiated by the data presented in Figure~\ref{fig:qint-experiment}b, which shows the parameters $\Gamma$ and $\Delta_\text{HF}$ as a function of $\theta_\text{D}$, fitted using either a pure Lorentzian model ($\mathcal{S}^{(b)}$, red data points), or equation \ref{eq:lineshapeFermions} ($\mathcal{S}^{(f)}$, blue data points). 
When a sum of two Lorentzians is used to model the lineshape, the linewidth and the measured hyperfine interval vary with the polarization angle. However, when equation \ref{eq:lineshapeFermions} is used, the polarization dependence disappears. The absolute position of the individual hyperfine components shifts by up to 20\,MHz, while the center of gravity shifts only by about 8\,MHz. We fit $A+BP_2(\cos\theta_D)$ to each of the data sets in Figure \ref{fig:qint-experiment}b). For the case of determining the hyperfine interval we get $B=25.4(3.2)$\,MHz (total error of 29(5)\,MHz) and for the case of the linewidth the fit yields $B=13.4(3.3)$\,MHz.
For the \TripletWL{} line, the maximum expected deviation $s_\text{Lor}(\Delta_\text{HF})\approx0.7$\,Hz. Measurements of the \Triplet{} transition are thus not sensitive to quantum interference on a level relevant to our study. The absolute frequencies are in this case independent of $\theta_D$, and we choose $\theta_D=\pi/2$ to maximize the detected boson fluorescence.

\subsubsection{Isotope-shift measurement}
For the measurements of the \Singlet{} transition, we record a spectrum of a ${}^{110}$Cd enriched sample and determine the absolute transition frequency before and after each spectrum with the mixed target (see Figure \ref{fig:spectra}). The fitted $^{110}$Cd line-centers of all five measurements agree within the standard errors of the fits and are thus averaged with a standard error of 0.4\,MHz. We then alternate between recording spectra of ${}^{110}$Cd and ${}^{112}$Cd enriched samples. Doing so allows us to precisely measure the IS(${}^{110,112}$Cd) as 457.5(7)\,MHz (statistical uncertainty) and to reduce the number of free parameters in the fit to the spectrum of Cd with natural abundance. The number of parameters is further reduced by fixing the relative amplitudes of each of the fermionic species in $\mathcal{S}^{(f)}$. We then fit a model that comprises a sum of eight terms, i.e., six Lorentzians ($\mathcal{S}^{(b)}$) for the bosons and two quantum interference line shapes ($\mathcal{S}^{(f)}$) for the fermions. This model fits the data well, as demonstrated by the small fit-residuals shown in Fig. \ref{fig:spectra}. The relative heights of the peaks are not fixed to the corresponding relative abundance in a natural sample since we expect source fluctuations of up to 10$\%$ over the course of a measurement. The deviation of the fitted relative abundance is consistent with this assumption. \\

\subsubsection{Radiative lifetime}

\begin{table}
\caption{\label{tab:lifetime} Comparison of the experimentally determined lifetime $\tau$ of the $^{1}\text{P}_1$ state in Cd with literature. Various indirect methods have been used for this task for nearly a century. To date, no direct measurement via the natural linewidth of the \Singlet{} transition has been reported.}
\begin{ruledtabular}
    \centering
    \begin{tabular}{lll}
     \text{$\tau$(ns)} &  \text{Year [Ref]} &  \text{Method} \\
     \hline
 2.00(8)    & \citeyear{Kuhn1926}~\cite{Kuhn1926} & magneto-rotation\footnotemark[1]\\
 1.99(10)   & \citeyear{Zemansky1931}~\cite{Zemansky1931} & line absorption\footnotemark[1]\\
 0.38       & \citeyear{Soleillet1943}~\cite{Soleillet1943} & magnetic depolarization\footnotemark[1]\\
 2.10       & \citeyear{Webb1944}~\cite{Webb1944}  & alternating voltage\footnotemark[2] \\
 1.70(9)    &  \citeyear{Welsh1950}~\cite{Welsh1950} & selective reflection\\
 1.66(5)    & \citeyear{Lurio1964}~\cite{Lurio1964} & Hanle-effect\\
 1.11       &  \citeyear{Spitzer1965}~\cite{Spitzer1965} & magnetic depolarization\footnotemark[1]\\
 1.66(5)    & \citeyear{Saussereau1969}~\cite{Saussereau1969}  & Hanle-effect\\
 2.1(3)     &  \citeyear{Baumann1970}~\cite{Baumann1970}  & phase-shift\\
 1.65(8)    & \citeyear{Pepperl1970}~\cite{Pepperl1970} & Hanle-effect\\
 1.90(15)   & \citeyear{Andersen1973}~\cite{Andersen1973} & beam-foil \\
 1.75(20)   & \citeyear{Xu2004}~\cite{Xu2004} & time-resolved fluorescence\\
 1.60(5)    & \textit{this work}  & lineshape\footnotemark[3]\\
       \end{tabular}
       \end{ruledtabular}
       
\footnotetext[1]{This method is today known as zero-field level crossing, also known as Hanle-effect}
\footnotetext[2]{Later referenced as phase-shift}
\footnotetext[3]{from $\tau=\Gamma^{-1}$ with \GammaMHz{} $= 99.7(3.3)$\,MHz}
\end{table}
The radiative lifetime of the $^1\text{P}_1$ state has been studied for almost a century. It has since been experimentally determined many times using various methods such as the Hanle-effect or time-resolved fluorescence (see Table \ref{tab:lifetime})\footnote{One of the first values was measured by \citeauthor{Zemansky1931}\cite{Zemansky1931} at the Kaiser-Wilhelm-Institut für physikalische Chemie und Elektrochemie, the original name of what currently is the Fritz Haber Institute of the Max Planck Society.}. Indirect methods used in the past to measure the lifetime of the $^1$P$_0$ state are more susceptible to systematic errors than a direct measurement of the lineshape.
Reported lifetimes range from 1.11-2.3\,ns with an outlier at 0.38\,ns. However, the authors themselves question this measurement. Nowadays, the values given by \citeauthor{Lurio1964}\cite{Lurio1964}\footnote{\citeauthor{Thaddeus1962}~\cite{Thaddeus1962} cite \citeauthor{Lurio1964}\cite{Lurio1964} as private communication with a slightly different value. In Table \ref{tab:lifetime} we use the published value.} (\GammaMHz{} $=95.9 \pm 2.9$\,MHz) and \citeauthor{Xu2004}\cite{Xu2004} (\GammaMHz{} $=91 \pm 10$\,MHz) are most commonly used in the literature.

To our knowledge, the lifetime has never been measured by fitting to a resonance line shape. Figure \ref{fig:qint-experiment}a) shows the spectrum of an enriched $^{113}$Cd ablation target. 
Using an enriched sample allows us to benchmark the model presented in equation \ref{eq:lineshapeFermions}, and extract the linewidth and the hyperfine interval from the fitted lineshape. In addition, we determine the lineshape and linewidth of the bosons by using an enriched target of $^{112}$Cd (not shown), where we use equation \ref{eq:lineshapeBosons} to fit the data. To improve the fit, we include the residual isotopes ($<2\%$), present in the enriched targets, in the fit. Doppler broadening is minimized by selecting atoms with a low forward velocity between $100-150$\,ms$^{-1}$ from the time-of-flight profile of the atomic beam. When atoms with a high forward velocity of $>300$\,ms$^{-1}$ are selected, the fitted linewidth remains unaffected. Fitting a Voigt profile for $^{112}$Cd does improve the fit residuals and results in a value for the Lorentzian contribution consistent with a regular Lorentzian model. In total, we record eight spectra of ${}^{112}$Cd and 11 spectra of ${}^{113}$Cd, including the 7 measurements presented in Figure \ref{fig:qint-experiment}. The spectral linewidths as extracted from the fits agree within the standard errors and are averaged to \GammaMHz{} $=99.7\pm0.6_{\text{stat}}\pm3.3_{\text{sys}}$\,MHz, which corresponds to $\tau(^1\text{P}_1)=1.60(5)$\,ns. The Lorentzian linewidth of the spectrum of an enriched ${}^{110}$Cd target is consistent within the 1\,MHz statistical error when applying a large magnetic field ($B=40$\,G or $4$\,mT) parallel to the laser polarization. Zeeman broadening induced by the uncancelled magnetic field in the detection region can thus be neglected. 

Experimentally obtained values for the lifetime of the $^3\text{P}_1$ state have been reported at least eight times with a weighted mean and standard error on the mean of 2.30(10)\,µs, corresponding to a linewidth of 69.1(27)\,kHz~\cite{byron1964,vandveer1990,schaefer1971,koenig1932,Webb1944,czajkowski1989,geneux1960,matland1953}. More recent literature refers to the value of \citeauthor{byron1964}~\cite{byron1964} which is 2.39(4)\,µs, corresponding to 66.6(11)\,kHz. Doppler broadening in our setup is two orders of magnitude larger than the natural linewidth, and we can thus not determine the lifetime by measuring the linewidth of the transition.

\subsection{Discussion}
\label{sec:discussion}
\begin{table}
 \caption{\label{tab:shift-comp}Isotope shifts relative to ${}^{114}$Cd, given in MHz. The systematic uncertainty to determine relative frequencies in this work is determined to 3.3\,MHz and should be added to the given statistical uncertainty. For fermionic isotopes, the quantum number of the excited state is given in brackets. For completeness, we include our results from Reference~\cite{Ohayon2022} and add the isotope shifts of the fermionic spectral components in bold.}%
\begin{ruledtabular}
  \begin{tabular}{lr@{}rr@{}r}
     isotope & \multicolumn{2}{c}{\SingletWL{}}& \multicolumn{2}{c}{\TripletWL{}}      \\
                & \textit{this work} & literature\footnotemark[1]  & \textit{this work} & literature\\
    \hline
     116        & -316.1(5)  & -299(4)  &   -326.9(2)   & -321.5(1.0)\footnotemark[2]  \\
     114        & 0.0(5)     & 0(4)     &   0.0(5)      & 0.0                          \\
     113(1/2)   & \textbf{147.8(4)}   & 148(4)   &   \textbf{4681.1(4)}   & 4653(19)\footnotemark[5]      \\
     &&&&4533(23)\footnotemark[3]\\
     112        & 407.5(7)   & 392(5)   &   426.3(3)    & 429.9(1.0)\footnotemark[2]   \\
     113(3/2)   & \textbf{443.4(7)}   & 427(5)   &   \textbf{-1785.2(3)}  & -1811(17)\footnotemark[3]     \\
     111(1/2)   & \textbf{616.5(5)}   & 592(6)   &   \textbf{4982.2(9)}   & 4947(24)\footnotemark[3]      \\
     110        & 865.0(3)   & 826(6)   &   909.3(6)    & 914.7(1.0)\footnotemark[2]   \\
     111(3/2)   & \textbf{899.2(4)}   & 875(6)   &   \textbf{-1205.3(4)}  & -1217(19)\footnotemark[3] \\
     108        & 1336.5(9)  & 1259(9)  &   1399.4(7)   & 1402.4(1.0)\footnotemark[2]  \\
     106        & 1818.1(1.4)& 1748(11) &   1911.2(2)   & 1913.0(1.0)\footnotemark[2]  \\
     111 c.g.   & \textbf{805.0(3)}   & 781(4)   &   \textbf{857.2(4)}    & 862(12)\footnotemark[4]    \\
     113 c.g.   & \textbf{344.9(5)}   & 334(4)   &   \textbf{370.2(2)}    & 374(11)\footnotemark[4]   \\
    \end{tabular}
\end{ruledtabular}
\footnotetext[1]{Beam measurement, \citeauthor{Tinsley2021}~\cite{Tinsley2021}}
\footnotetext[2]{MOT edge measurement, \citeauthor{Ohayon2022}~\cite{Ohayon2022}}
\footnotetext[3]{Beam measurement, \citeauthor{Kelly1959}~\cite{Kelly1959}}
\footnotetext[4]{Hollow cathode discharge, relative to $^{106,108}$Cd, \citeauthor{Kloch1987}~\cite{Kloch1987}}
\footnotetext[5]{ \citeauthor{hanes1955}~\cite{hanes1955}}
\end{table}

Table \ref{tab:shift-comp} summarizes our measured ISs relative to the ${}^{114}$Cd line and compares them to  literature values. The uncertainty given is the standard error of the mean of several measurements. The values for the ISs are consistent with a recent measurement using a magneto-optical trap of Cd~\cite{Ohayon2022}. For the \Singlet{} transition, our results are more precise than a recent measurement~\cite{Tinsley2021}, and we observe a large discrepancy that increases linearly with the frequency. Our results for the \Triplet{} transition agree well with other literature (see overview in~\cite{Maslowski2009}) and are significantly more precise and accurate, especially for the fermions. 
\subsubsection{King plot}
\begin{figure}
  \caption{\label{fig:king-plot}King-plot analysis of the ISs of \SingletWL{} and \TripletWL{} relative to $^{114}$Cd. The error bars include the systematic uncertainty in determining relative frequencies of 3.3\,MHz. The shaded areas show the 68$\%$ confidence intervals for a linear fit to the bosons only (light blue). The fit including the fermions is shown in dark blue. The ISs of the fermions (center of gravity) are consistent with the extrapolated linear fit to the bosons. The inset shows a zoom-in.}
  \includegraphics[]{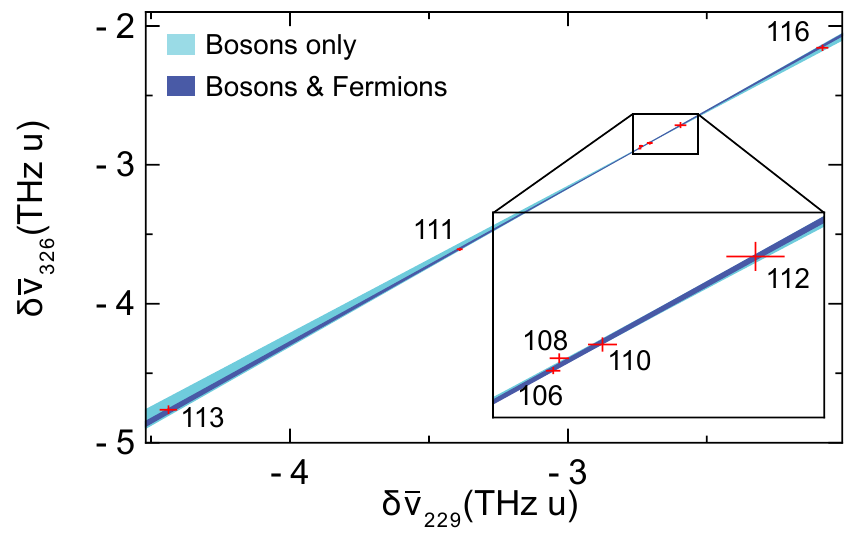}
\end{figure}
An accurate set of ISs on two transitions can be analyzed in a King plot. Figure \ref{fig:king-plot} shows such a plot using the modified IS $\delta \bar \nu_i^{A,A'}=\delta \nu_i^{A,A'}/\mu^{A,A'}$, where $\mu^{A,A'}=1/M_{A}-1/M_{A'}$ is the nuclear mass-shift term. The linear fit has the form: 
\begin{align*}
    \delta \bar{\nu}_i^{A,A'}=F_{ij}\delta\bar{\nu}_j^{A,A'}+K_{ij},
\end{align*}
where $F$ and $K$ are the FS and MS coefficients, respectively, $F_{ij}=F_i/F_j$ is the slope and $K_{ij}=K_i-F_{ij}K_j$ is the y-axis intercept. The fit results in $F_{ij}=1.10(3)$ and $K_{ij}=0.13(10)$\,THz~u, if only the bosons are considered. Including the fermions reduces the fitting uncertainty by a factor of $\approx 3$ to
\begin{align*}
    F_{ij}=1.117(12)\text{ and } K_{ij}=0.18(3)\,\text{THz\,u.}
\end{align*}
The fact that including the fermions reduces the uncertainty of the fit shows that potential shifts due to the second-order hyperfine interaction are negligible on the MHz-level.
These results can be compared to CI-MBPT calculations (from preprint~\cite{Han2021}), which yield $F_{ij}=1.133(80)$ and $K_{ij}=0.25(49)$\,THz~u. These results are consistent with ours, but our uncertainty is much lower, setting stringent benchmarks to improve the calculation. We note that the uncertainty of the calculated field shift ratio $F_{ij}$ is probably much smaller than our estimate from literature, due to correlations.

\subsubsection{Nuclear charge radii}
Our ISs for the fermions, measured using a narrow line where nearby peaks do not overlap and quantum interference is negligible, combined with the calculated negligible off-diagonal hyperfine shifts for this line~\cite{Schelfhout2022}, enable a reliable extraction for the RMS charge radius difference $(\delta r^2)^{A,114}$ from the center of gravity ISs given in Table \ref{tab:shift-comp}, provided that the atomic parameters for this line are known.
Here we use $F_{326}=-4354(62)$\,MHz\,fm$^{-2}$ and $K_{326}=1673(43)$\,GHz\,u obtained by projecting accurate calculations in Cd$^+$ using a King plot containing only bosons~\cite{Ohayon2022}. 
$(\delta r^2)^{A,114}$ are given in Table \ref{tab:nuclear charge radii}. 
They compare reasonably well to within two combined standard deviations with those obtained through muonic x-ray measurement~\cite{2004-FrickeCd}, a calibrated King plot combining muonic x-rays and ISs in Cd$^+$~\cite{2022-Wang, Hammen2018}, and a determination based on direct calculation of $F_{326}$ and $K_{326}$~\cite{Schelfhout2022}.
The slight deviation from~\cite{Schelfhout2022} is ascribed to the fact that the calculation only reports numerical uncertainties. The slight deviation from determinations using muonic atoms is most likely due to their dependency on extrapolations of unknown higher-moment corrections from electron scattering, which has not been performed for most Cd isotopes.

 \begin{table}
 \caption{\label{tab:nuclear charge radii}Nuclear charge radius difference $(\delta r^2)^{A,114}$ in fm$^2$ extracted from Table \ref{tab:shift-comp} with F$_i$ and K$_i$ from Reference~\cite{Ohayon2022}.}%
\begin{ruledtabular}\begin{tabular}{ccccc}
     isotope & \textit{this work} & Ref.~\cite{Schelfhout2022} & Ref.~\cite{2004-FrickeCd}& Ref.~\cite{2022-Wang}\\
    \hline
      111 & -0.296(5)   & -0.269(15) & -0.289(4)\footnotemark[1] & -0.285(4)\\
      113 & -0.118(2) & -0.101(10) & -0.116(4)\footnotemark[1] & -0.113(2)\\
    \end{tabular}
\end{ruledtabular}
\footnotetext[1]{Our estimate is based on the reported Barret radii.}
\end{table}
\subsubsection{Hyperfine intervals}
Table \ref{tab:hfsshift} compares our measured hyperfine intervals for the $^1$P$_1$ and $^3$P$_1$ state with literature. For the $^1$P$_1$ state, we show that the spectral overlap caused a significant systematic error in previous experiments.
For the $^3$P$_1$ state, we find that our values agree well with precise double-resonance measurements\cite{Lacey1952thesis, Lacey1959,Thaddeus1962}. Recent atomic structure calculations~\cite{Schelfhout2022,lu2022} agree to some extent and can be benchmarked with our results. This also allows us to benchmark the systematic uncertainty in the linearity of our wavemeter, as shown in Section\,\ref{sec:systematic}. 

\begin{table}
 \caption{\label{tab:hfsshift}Comparison of experimentally determined values for the hyperfine splitting $\Delta_\text{HF}$(${}^{111}$Cd, ${}^{113}$Cd) in MHz with literature values.}%
\begin{ruledtabular}
  \begin{tabular}{lr@{}rr@{}r}
      &\multicolumn{2}{c}{\SingletWL{}}& \multicolumn{2}{c}{\TripletWL{}}      \\
         &${}^{111}$Cd& ${}^{113}$Cd & ${}^{111}$Cd& ${}^{113}$Cd\\
    \hline
    \textit{this work}                 &  282.7(3.3)   &   295.6(3.4)  &   6187.5(3.3)     &   6466.3(3.4) \\
    \cite{Tinsley2021},\cite{Thaddeus1962} &   285(7)      &   251(5)      &   6185.72(2)  &   6470.79(2)  \\
    \cite{Maslowski2009}                &   -           &   -           &   -           &   6444(18)    \\
    \cite{Kelly1959}                    &   -           &   -           &   6164(31)    &   6344(28)    \\
    \cite{hanes1955}                    &   -           &   -           &   6183(30)    &   6465(21)    \\
    \end{tabular}
\end{ruledtabular}
\end{table}
\subsubsection{Absolute frequencies}
The fitted line centers of spectra taken over a period of a few weeks agree with the uncertainties of the fits.
Using a calibrated wavemeter allows us to determine the absolute transition frequencies of the ${}^{114}$Cd isotope for the two transitions:
\begin{align*}
\begin{tabular}{rrr}
${}^{114}$Cd (\Singlet{}):& 1,309,864,341(20)   & MHz\\
${}^{114}$Cd (\Triplet{}):& 919,046,234(21)     & MHz
\end{tabular}
\end{align*} 
The absolute frequency of the \Singlet{} transition is consistent with a recent measurement  (1,309,864,506(262)\,THz~\cite{Tinsley2021}) and 13 times more precise. The value given in reference~\cite{Burns1956} (1,309,864,580(86)\,THz) disagrees by about three standard deviations. The same reference~\cite{Burns1956} also lists the absolute frequency for the \Triplet{} transition to  919,046,357(42)\,THz, which is also higher by about three times the stated standard deviation.

\subsection{\label{sec:systematic}Systematic uncertainty}
Our measurements have a typical statistical uncertainty of $\leq 1$\,MHz, which means we can reliably determine the line center of features in the \Singlet{} transition to $\Gamma/100$. However, the wavemeter we use as a frequency reference is not necessarily linear over a range of several GHz. We anticipate that this will be the limiting factor in the accuracy when determining relative frequencies. The hyperfine intervals of the $^3$P$_1$ state are known with high accuracy and already provide a good indication (see Table \ref{tab:hfsshift}). Furthermore, the ratio of the nuclear magnetic moments of Cd determines the ratio of the hyperfine intervals for different isotopes and is known with high accuracy. We perform two additional tests to estimate this uncertainty.
First, we measure the hyperfine structure of $^{63,65}$Cu near 327.5\,nm and 324.8\,nm, which span a frequency range of up to 12.6\,GHz and are known with a precision of 10\,kHz.
Second, we compare the linearity of our wavemeter to an ultra-stable cavity. The systematic uncertainty in determining relative frequencies is estimated conservatively to be 3.3\,MHz. This uncertainty is independent of the frequency span and assumed to be identical for all measurements. For absolute frequency measurements, we are limited by the absolute accuracy of the wavemeter, which is 10\,MHz in the fundamental.

\subsubsection{\label{sec:hyperfinesplitting}\Triplet{} nuclear magnetic moment ratios}
The measured hyperfine splittings at \TripletWL{} are given in Table \ref{tab:hfsshift}. We find good agreement with precise literature values. Furthermore, the ratio of the nuclear magnetic moments g($^{113}$Cd)/g($^{111}$Cd) has been determined with high accuracy to 1.0460842(2)\cite{SPENCE1972}. In the approximation that hyperfine anomaly is negligible, we can assume that g($^{113}$Cd)/g($^{111}$Cd)=$\Delta_\text{HF}$($^{113}$Cd)/$\Delta_\text{HF}$($^{111}$Cd). For the \Singlet{} transition we use this ratio as a fixed parameter.
In our measurements of the \Triplet{} transition, we find a ratio of 1.04506(18) (statistical), which indicates an upper bound for the hyperfine anomaly of 4.3\,MHz.

\subsubsection{Hyperfine structure of Copper}
\begin{figure}
    \centering
    \includegraphics[]{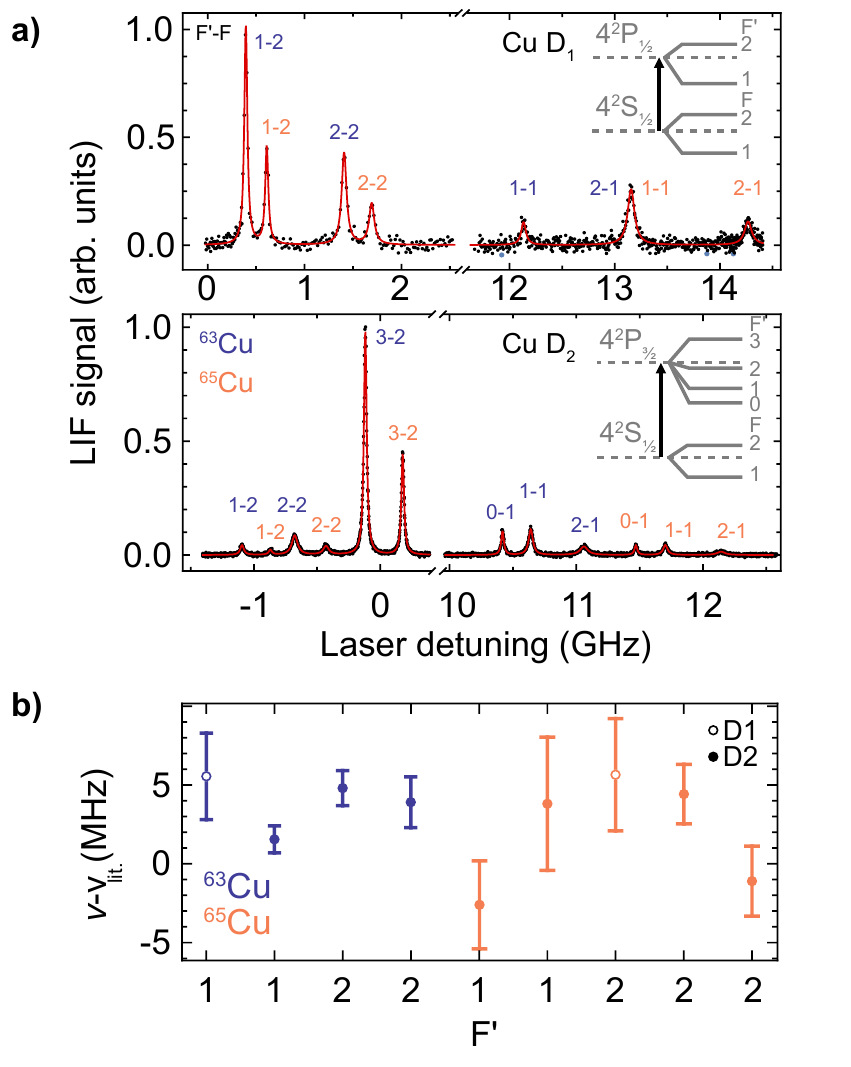}
    \caption{Copper spectra of the D$_1$ and D$_2$ lines near $326$\,nm to benchmark the linearity of our wavemeter. a) Laser-induced fluorescence spectra of hyperfine transitions. Experimental data points in black and a Lorentzian fit (red curve). b) Comparison of the measured hyperfine intervals measured via a common excited state $F'$ to kHz-accuracy microwave data from \citeauthor{Ting1957}~\cite{Ting1957}.}
    \label{fig:copper}
\end{figure}

The D$_1$ (327.5\,nm) and D$_2$ (324.8\,nm) lines of Cu I lie about 1.3\,nm on either side of the Cd \TripletWL{} transition. The hyperfine splitting of the ground state is known with an accuracy of 10\,kHz and has been determined to 12568.81(1)\,MHz for $^{65}$Cu and to 11733.83(1)\,MHz for $^{63}$Cu~\cite{Ting1957}. For this, we only replace the Cd ablation target in our buffer gas source with a Cu one and leave the remaining setup identical. We measure the ground state hyperfine splitting by the method of combination differences, using transition pairs involving a common excited state with total angular momentum $F'$. Spectra of the D$_1$ and D$_2$ lines are shown in Figure \ref{fig:copper}a). The difference between the extracted ground state splitting with the precisely known values are shown organized by isotope and $F'$ in Figure \ref{fig:copper}b). The mean deviation of these frequency differences is 3.1$\pm$2.0\,MHz for $^{63}$Cu and 1.8$\pm$3.7\,MHz for $^{65}$Cu. The weighted mean and standard error of the mean of all measurements indicates a systematic error of 3.1(1.9)\,MHz.

\subsubsection{Stable Fabry-Pérot Cavity}
\begin{figure}
    \centering
    \includegraphics[]{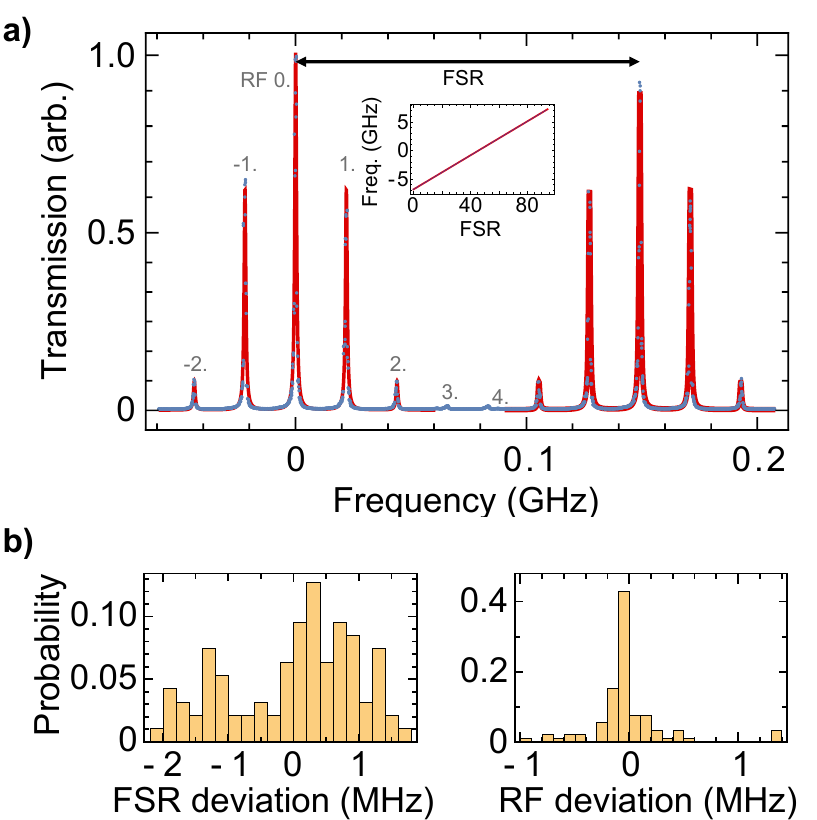}
    \caption{\label{fig:wavemeter}Comparison of wavemeter linearity to a stable Fabry-P\'{e}rot cavity. a) Transmission through the cavity (free spectral range of about 150\,MHz) at $740$\,nm as a function of the laser frequency. An electro-optic modulator phase modulates the laser with a fixed RF frequency. The inset compares the frequency as read by the wavemeter to the number of FSR peaks. b) The probability distribution of the residuals of a linear fit to the data has a standard deviation of 1.0\,MHz. The observed spacing of the RF sidebands as measured by the wavemeter has a standard deviation of 0.34\,MHz.}
\end{figure}

 To measure the linearity of our wavemeter with better statistics than in the previous sub-sections, we use a stable cavity in combination with an electro-optic modulator (EOM) that phase-modulates the laser to generate radio-frequency sidebands spaced by precisely 22.051\,MHz. The cavity is a near-confocal, pressure and temperature-stabilized Fabry-P\'{e}rot cavity with a Zerodur spacer, a free spectral range (FSR) of 150\,MHz and a drift rate of $7.5$\,Hz per FSR per Kelvin\cite{bardizza2005}. The high-reflectivity coating of the cavity mirrors drops sharply above $740$\,nm and we therefore tune the Ti:Sa to $740$\,nm. Scanning the laser while monitoring the cavity transmission on a photodiode, as shown in Figure \ref{fig:wavemeter}a), produces a comb of resonance peaks with a known frequency spacing. By fitting the transmission peaks on the frequency axis obtained with the wavemeter, the short-term (minutes) and long-term (hours) linearity of the wavemeter can be determined. Figure \ref{fig:wavemeter}b) shows that the wavemeter is close to linear for short-term scans over a single FSR and long-term when scanning slowly over a frequency range of about 14\,GHz. Fitting to the cavity peaks gives a standard deviation of 1.0\,MHz from the FSR for long-term scans (see inset in Figure \ref{fig:wavemeter}a) and 0.34\,MHz from the RF sidebands for short-term scans, which is close to the wavemeter resolution of 0.4\,MHz.
 We repeat this measurement with the ring dye laser at 652\,nm. From several measurements on different days using both spectroscopy lasers, we determine a conservative upper limit for the systematic uncertainty in the determination of the relative frequencies at \SingletWL{} and \TripletWL{} of 3.3\,MHz.
\subsubsection{Absolute accuracy}
We use a temperature-stabilized HeNe laser at $633$\,nm whose absolute frequency is calibrated to 5\,MHz as a reference for the wavemeter. Recently, we also measured the ISs of Yb to benchmark the absolute calibration and linearity of our wavemeter near 399\,nm~\cite{Doppelbauer2022}. The known absolute frequencies in Yb were reproduced with an accuracy of 6\,MHz and the relative ISs were measured with a standard deviation of 1.3\,MHz. For the measurements presented here, we couple the frequency-doubled light of the Ti:Sa into the wavemeter to avoid the effect of intermittent multimode frequency content at the fundamental wavelength. For the \TripletWL{} light we record the fundamental wavelength of the dye laser since fiber transmission at \TripletWL{} is limited and influenced over time by photodegradation. We determine the absolute frequency of ${}^{114}$Cd from a weighted average of all measurements performed. The absolute frequency for \SingletWL{} was repeatedly found with less than 1\,MHz standard deviation. For \TripletWL{} it was reproducible over a period of nine weeks with a maximum spread of 12\,MHz and a standard deviation of 5\,MHz. The manufacturer specifies the absolute accuracy in the fundamental light to be 10\,MHz, which we take as an upper bound for the uncertainty in determining absolute frequencies.

\section{\label{sec:conclusion}Summary}
We measured the isotope shift in Cd-I for the \Singlet{} and \Triplet{} transitions with a statistical uncertainty of typically $<1$\,MHz by fitting to the laser-induced fluorescence spectra from a pulsed cryogenic buffer gas cooled atomic beam. Measurements of the \Singlet{} transition are complicated by significant overlap of the spectral lines. We resolve this issue by using enriched Cd ablation targets and by suppressing the emission from the bosonic isotopes towards the direction of the detector. We determined the hyperfine intervals in the $^1$P$_1$ state of the fermionic isotopes and showed that quantum interference effects affect the spectral lineshape.
The accuracy of our wavemeter is calibrated against well-known hyperfine intervals in Cd and Cu and by using a stable reference cavity in combination with an EOM. We find an upper bound of 3.3\,MHz for the systematic uncertainty in measuring relative shifts. The lifetime of the $^1$P$_1$ state is determined spectroscopically to be 1.60(5)\,ns, and the absolute transition frequencies were determined with unprecedented accuracy. Our results differ significantly from recent measurements, demonstrating the importance of understanding the spectral lineshape and measuring the linearity of the spectrometer. A King plot comprising both transitions was presented. All naturally occurring Cd isotopes follow a linear relation and the fitted slope and intercept are consistent with recent atomic structure calculations. The second-order hyperfine interaction in the fermions is negligible at the MHz level. Combining our new measurements with recent calculations of the isotope-shift parameters allowed us to extract the fermionic isotopes' precise nuclear charge radius differences. The measurements presented here resolve significant discrepancies in the recent literature and  benchmark new atomic structure calculations; a first important step towards using King plots of Cd to search for new physics beyond the Standard Model. A large number of naturally occurring isotopes, the presence of narrow optical transitions, and the expected small Standard Model background make Cd an ideal candidate for such searches.   

\section*{Acknowledgements}
This project received funding from the European Research Council (ERC) under the European Union’s Horizon 2020 Research and Innovation Programme (CoMoFun, Grant Agreement No. 949119).

\bibliography{APSversion}

\end{document}